\documentclass[%
 reprint,
 amsmath,amssymb,
 aps,
]{revtex4-1}

\usepackage{graphicx}% Include figure files
\usepackage{dcolumn}% Align table columns on decimal point
\usepackage{bm}% bold math
\usepackage{xcolor}
\begin{document}

\preprint{APS/123-QED}

\title{Spintronic signatures of Klein tunneling in topological insulators}% Force line breaks with \\
%\thanks{A footnote to the article title}%

\author{Yunkun Xie}
\email{yx3ga@virginia.edu}
 \altaffiliation{Charles L. Brown Department of Electrical and Computer Engineering, University of Virginia, Charlottesville, VA, 22904 USA.}%Lines break automatically or can be forced with \\
 
\author{Yaohua Tan}%
% \email{Second.Author@institution.edu}
\affiliation{%
 Charles L. Brown Department of Electrical and Computer Engineering, University of Virginia, Charlottesville, VA, 22904 USA.
}%

\author{Avik W. Ghosh}
\affiliation{Charles L. Brown Department of Electrical and Computer Engineering, University of Virginia, Charlottesville, VA, 22904 USA.
}%

\date{\today}% It is always \today, today,
             %  but any date may be explicitly specified

\begin{abstract}
Klein tunneling, the perfect transmission of normally incident Dirac electrons across a potential barrier, has been widely studied in graphene and explored to design switches, albeit indirectly. We show that Klein tunneling maybe easier to detect for spin-momentum locked electrons crossing a PN junction along a three dimensional topological insulator surface. In these topological insulator PN junctions (TIPNJs), the spin texture and momentum distribution of transmitted electrons can be measured electrically using a ferromagnetic probe for varying gate voltages and angles of current injection. Based on transport models across a TIPNJ, we show that the asymmetry in the potentiometric signal between PP and PN junctions and its overall angular dependence serve as a direct signature of Klein tunneling. 
\end{abstract}

\pacs{Valid PACS appear here}% PACS, the Physics and Astronomy
                             % Classification Scheme.
%Use showkeys class option if keyword
                              %display desired
\maketitle

%\tableofcontents

%\section{Introduction}
The surface of 3D topological insulators (TIs) such as Bi$_2$Se$_3$ has a simple Dirac  cone band structure\cite{chen2009experimental} reminiscent of graphene, except its branches are labeled by spins rather than pseudospins. Carriers along the surface have their spins locked with their linear momentum \cite{qi2011topological}, which can generate polarized spins with charge injection and apply a sizeable spin torque on a magnet \cite{mellnik2014spin,han2017room,jamali2017room}. Recently we suggested that a TIPNJ can be used as a gate tunable spin filter to amplify charge to spin conversion at a magnetic source and increase spin polarization at the drain \cite{habib2015chiral}. Such a tunable torque can have potential applications in all spin logic \cite{behin2010proposal}. Beyond applications, the TI surface state offers opportunities to study the fundamental physics of Dirac electrons such as Veselago focusing and Klein tunneling \cite{klein1929reflexion}. Klein tunneling has been widely studied in graphene, albeit indirectly. It has been invoked to engineer a gate tunable pseudogap in graphene at high mobility, making it potentially useful for both low power digital and high speed analog switches \cite{beenakker2008colloquium,stander2009evidence,sajjad2013manipulating,chen2016electron}. While all these studies probed the charge current, the TI surface provides an appealing simple alternative to analyze momentum collimation, namely, by directly monitoring the angle-dependent transmission of the electron spins using a ferromagnetic tip. 

It has been proposed\cite{hong2012modeling,sayed2016multi} and experimentally demonstrated that a potentiometric measurement with a ferromagnetic probe could measure the polarization of TI surface spins\cite{li2014electrical,lee2015mapping,tian2015electrical,li2016electrical}. In this paper, we extend this idea to a TIPNJ and demonstrate from detailed calculations that the angle and voltage dependent potentials measured at the probe bear direct signatures of Klein tunneling across the PN junction.

\begin{figure}
\includegraphics[width=8cm]{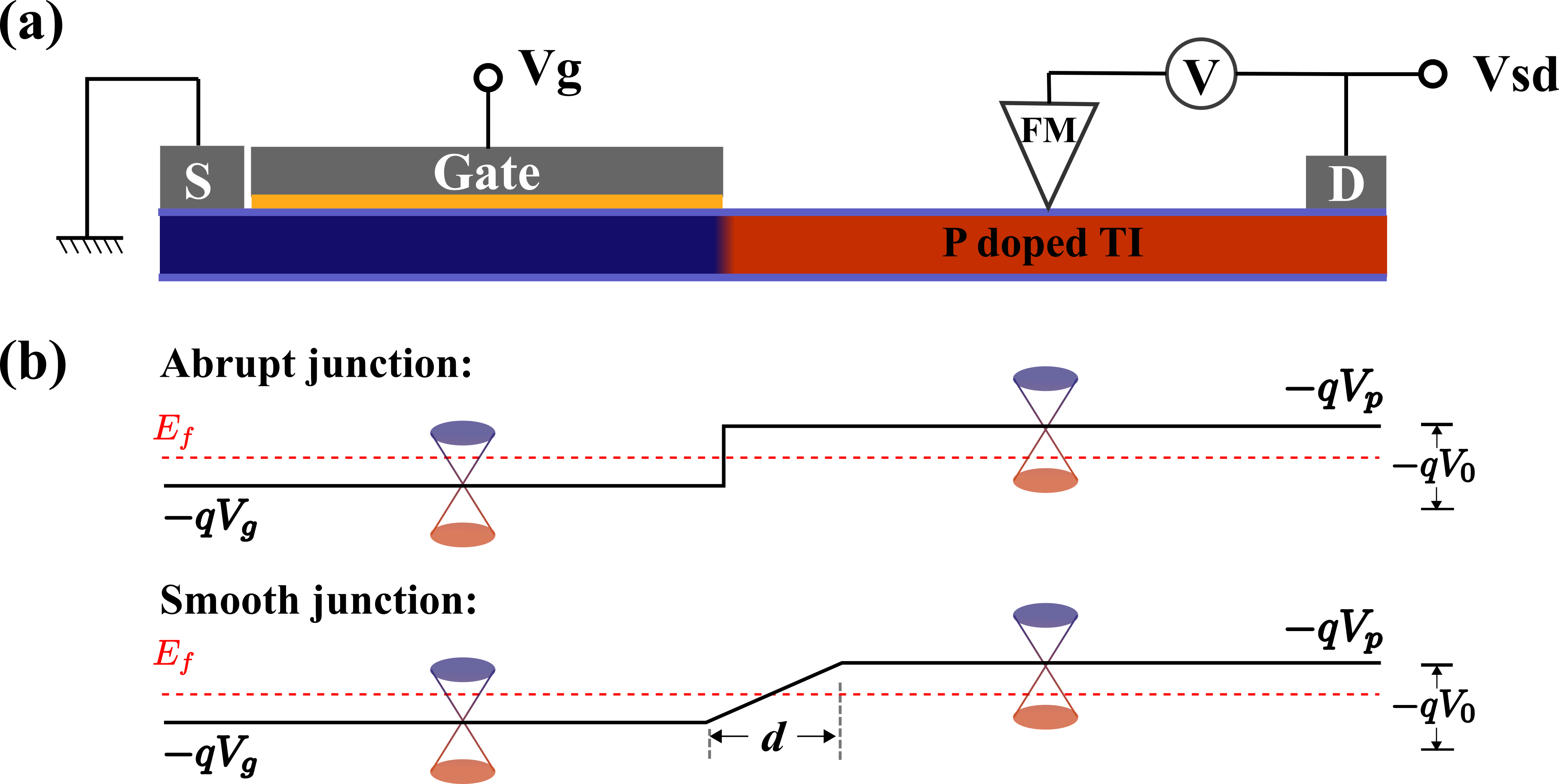}
\caption{(a) A basic setup for potentiometric measurement on a topological insulator PN junction. (b) The electrostatic potential profiles (abrupt of smooth) across TI PN junction.}
\label{fig:TI_struc}
\end{figure}

%\section{\label{sec:level1}Transport modeling in TIPNJ}
Fig.~\ref{fig:TI_struc}(a) shows a schematic structure of the TI pn junction in a potentiometric measurement setup. The TI surface can be chemically doped into P or N-type, as demonstrated in multiple experiments \cite{zhou2012controlling,Tu2016}. The figure shows a P-doped TI surface with a top gate on the source side that can swing it electrostatically to N-type. The rest of the P-type TI surface is exposed and a ferromagnetic probe is placed on top of the exposed surface to monitor the voltage at different gate bias and angular orientations (the orientation can be altered by using multiple contacts at relative angles, as we discuss later).

The TI surface states can be described by the $k\cdot p$ Hamiltonian when the electron energy under consideration is close to the Dirac point\cite{qi2011topological}: 
\begin{equation}
H=v_F\mathbf{\hat{z}}\cdot\left(\boldsymbol{\sigma}\times \mathbf{p}\right)
\label{eq:TI_H}
\end{equation}
where $\mathbf{\hat{z}}$ is the normal vector of the surface and $v_F$ is the speed of electrons near the Dirac point. $\boldsymbol{\sigma}=(\sigma_x,\sigma_y,\sigma_z)$ are the Pauli matrices. It should be emphasized that this parameterized surface Hamiltonian ignores any bulk leakage current that could control the strength of the measured voltage. In binary TI compounds such as $\mathrm{BiSb,\,Bi_2Se_3,\,Bi_2Te_3}$, it can be challenging to separate the surface contribution from the dominant bulk contribution\cite{checkelsky2009quantum,taskin2009quantum,butch2010strong}. One possible solution is to use ternary compounds like $\mathrm{Bi_2Te_2Se}$ with low carrier density in the bulk \cite{jia2011low}. Minimizing the leakage current into the bulk of TI is still an active research topic that is outside the scope of this paper. Here we only discuss the pure surface states of 3D TI.

The electrostatic potential across the TI PN junction is given by:
\begin{equation}
\begin{split}
V(x)&=-qV_p,\;\;\text{exposed P side}\\
&=-qV_g,\;\;\text{gate side}
\end{split}
\end{equation}
where $E_p=-qV_p$ is the energy difference between the local electron chemical potential and the Dirac point ($E=0$). $V_g$ is the gate voltage on the source side as shown in Fig.\ref{fig:TI_struc}(a). Two potential profiles are depicted in Fig. \ref{fig:TI_struc}(b), one with an abrupt potential change at the junction interface while the other assumes a smooth transition. We first derive the equations  based on an abrupt junction (Appendix \ref{sec:analytical}) and then extend it to a smooth junction, which is closer to a realistic profile\cite{gutierrez2016klein}. For smooth junctions, the transition region between N and P is set to $50\,\mathrm{nm}$ wide and the FM probe is placed $80\,\mathrm{nm}$ from the junction interface.

{\it{Formalism.}} In the ballistic limit, the electrons only scatter at the PN junction. A weakly coupled ferromagnetic voltage probe can detect the local chemical potential of the non-equilibrium electrons with different spin orientations. To calculate the voltage measured by the FM probe, we treat it as a third contact (B\"uttiker probe) besides source and drain. From Landauer theory \cite{datta1997electronic,ghosh2016nanoelectronics}, we can estimate that the voltage probe exchanges electrons with the TI surface through the following equations:
\begin{equation}
\begin{split}
I_{in}&=\mathrm{Tr}\left[\Gamma_\mathrm{FM}G^n\right]=\mathrm{Tr}\left[\Gamma_\mathrm{FM}\left(f_sA_s+f_dA_d\right)\right]\\
I_{out}&=f_p\mathrm{Tr}\left[\Gamma_\mathrm{FM}A\right]=f_p\mathrm{Tr}\left[\Gamma_\mathrm{FM}\left(A_s+A_d\right)\right]
\end{split}
\label{eq:NEGF}
\end{equation}
where $I_{in}\,(I_{out})$ is the incoming (outgoing) currents through the probe. $\Gamma_\mathrm{FM}$ is the coupling between the FM probe and the TI surface. $G^n$ is the correlation matrix while $A_s(A_d)$ are the partial spectral functions populated by the source (drain). $A=A_s+A_d$ is the total spectral function. $f_s$, $f_d$, $f_p$ are the Fermi-Dirac distribution functions of the source, drain and the floating probe respectively. In general, the spectral functions can be calculated numerically through the NEGF formalism (see appendix  \ref{sec:numerical}):

\begin{eqnarray}
A_s = G^R\Gamma_sG^{R\dagger},\quad \Gamma_s=i(\Sigma_s-\Sigma_s^{\dagger})\nonumber\\
A_d = G^R\Gamma_dG^{R\dagger},\quad \Gamma_d=i(\Sigma_d-\Sigma_d^{\dagger})
\end{eqnarray}
where $G^R$ is the retarded green's function and $\Sigma_{s,d}$ are self-energies for the source and the drain. In our simple setup, quasi-analytical results for $A_{s,d}$ are worked out in the appendix \ref{sec:analytical}.

The coupling between the FM probe and the TI surface depends on the magnetization of the FM probe $\mathbf{m}=(m_x,m_y,m_z)$ and electron spin on the TI surface:
\begin{equation}
\Gamma_\mathrm{FM}(\mathbf{m})=\gamma_0\left(1+P_\mathrm{FM}\mathbf{m}\cdot\boldsymbol{\sigma}\right)
\label{eq:couple}
\end{equation}
where $\gamma_0=\frac{\gamma_p+\gamma_{ap}}{2}$ is the average coupling between the FM probe and the TI surface when the magnetization of the probe is in parallel or anti-parallel alignment with the surface electron spin. $P_\mathrm{FM}=(\gamma_p-\gamma_{ap})/(\gamma_p+\gamma_{ap})$ is the `polarization' of the FM probe, representing the sensitivity of the FM probe to the electron spins. 

The voltage signal measured by the FM probe is determined by its distribution function $f_p$, which can be solved based on the condition that a voltage probe draws zero net current $I_{in}=I_{out}$:

\begin{eqnarray}
f_p(\mathbf{m}) &=& \frac{ \left(f_s-f_d\right)\mathrm{Tr}\left[\Gamma_\mathrm{FM}A_s\right]}{\mathrm{Tr}\left[\Gamma_\mathrm{FM}A\right]}+f_d\nonumber\\
&=&\lambda(\mathbf{m})(f_s-f_d) + f_d 
\label{eq:fp}
\end{eqnarray}
$f_p$ varies when the magnetization $\mathbf{m}$ points to different directions. We use the dimensionless parameter $\lambda(\mathbf{m})$ to characterize the dependence of the voltage signal on the direction of the magnetization. At low-temperature and small bias, the Fermi-Dirac distribution reduces to a step function and chemical potential of the probe can be expressed as:
\begin{eqnarray}
\mu_p(\mathbf{m}) = \lambda(\mathbf{m})(\mu_s-\mu_d) + \mu_d 
\label{eq:mup}
\end{eqnarray}
Experimentally instead of switching the magnetization of the FM probe we can drive current along two opposite directions (source to drain and vice-versa), then relate the measured voltage difference $\mu_p(\mathbf{m})-\mu_p(-\mathbf{m})$ to $\Delta\lambda(\mathbf{m})=\lambda(\mathbf{m})-\lambda(\mathbf{-m})$ through the charge current and the ballistic resistance of the junction:
\begin{eqnarray}
\Delta\lambda(\mathbf{m})&=&\frac{\mu_p(\mathbf{m})-\mu_p(\mathbf{-m})}{qIR_B} \nonumber\\
R_B&=&\frac{h}{q^2T(E_f)}
\label{eq:dlamdba}
\end{eqnarray}
where $R_B$ is the gate voltage dependent ballistic resistance of the junction, calculated using the average transmission at the fermi energy from Eq. \ref{eq:charge_current}.

We can further define a quantity $p(\mathbf{m})$ for the measured `polarization' of the TI surface electrons along the magnetization direction $\mathbf{m}$:
\begin{eqnarray}
p(\mathbf{m}) &=& \frac{\lambda(\mathbf{m})-\lambda(\mathbf{-m})}{\lambda(\mathbf{m})+\lambda(\mathbf{-m})}\nonumber\\
&=&\frac{\mu_p(\mathbf{m})-\mu_p(\mathbf{-m})}{\mu_p(\mathbf{m})+\mu_p(\mathbf{-m})-2\mu_d} 
\label{eq:polarization_TI}
\end{eqnarray}
The physical interpretation of Eq. \ref{eq:polarization_TI} becomes obvious when we substitute Eq. \ref{eq:couple} into Eq. \ref{eq:polarization_TI} and see that $\mathrm{Tr}\left[\Gamma_\mathrm{FM}(\mathbf{m})A\right]=\mathrm{Tr}\left[\Gamma_\mathrm{FM}(\mathbf{-m})A\right]$ due to the time reversal symmetry of TI surface states (see Eq.\ref{eq:As_analy}). Eq. \ref{eq:polarization_TI} reduces to:
\begin{eqnarray}
p(\mathbf{m}) = P_\mathrm{FM}\frac{\mathrm{Tr}[(\mathbf{m}\cdot\boldsymbol{\sigma})\gamma_0A_s]}{\mathrm{Tr}[\gamma_0A_s]} 
\label{eq:polarization_TI2}
\end{eqnarray}
when Eq. \ref{eq:polarization_TI2} is evaluated in the bias window, it indicates the spin polarization of the non-equilibrium electrons along direction $\mathbf{m}$. Notice that $P_\mathrm{FM}$ also appears in the equation to account for the sensitivity of FM probe. Our definition is compatible with the polarization defined in \cite{hong2012modeling} for homogeneous TI surface.

\begin{figure}
\includegraphics[width=8cm]{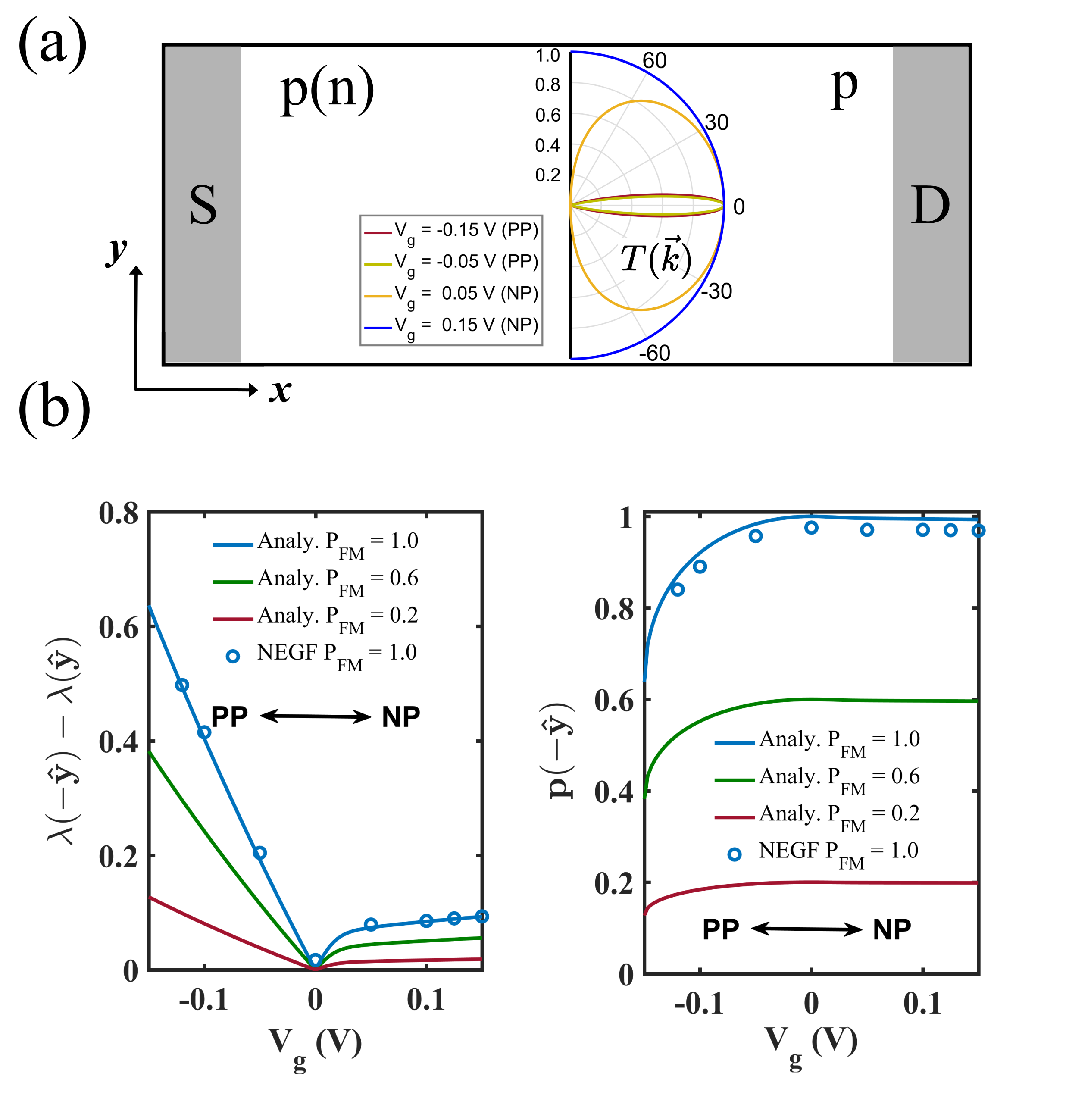}
\caption{(a) Schematic plot of the electron transmission through the junction at different gate voltages. (b) Left. Gate voltage dependence of $\Delta\lambda(-\mathbf{\hat{y}})$ for various probe sensitivities. (b) Right. The measurable polarization of TI surface electrons along $\hat{y}$ direction. The circles are benchmark results from NEGF simulations.}
\label{fig:smooth_pn}
\end{figure}

{\it{Varying gate voltage: from PP to NP junction.}}
\label{sec:vg_dep}
The impact of a TIPNJ on surface electron transport is worked out in appendix \ref{sec:analytical} and summarized schematically in Fig. \ref{fig:smooth_pn}(a).  Consider a small source-drain bias near the Fermi energy, as shown in Fig. \ref{fig:TI_struc}(b). As the gate voltage varies from $V_g=V_p$ to $V_g=-V_p$, the TI switches from a homogeneous P-doped surface to an NP junction. Electrons see a potential barrier from the N region to the P region. In a normal semiconductor, such a barrier creates decaying electron waves in the P region and results in a vanishing current. For Dirac type TI surface, however, the junction acts like a collimator for electrons, filtering out electrons with large incident angles but preserving the normally incident modes that cannot back-scatter due to spin conservation. The resulting electron transmission for various gate voltages is plotted in Fig. \ref{fig:smooth_pn}(a). This behavior can translate to the gate voltage dependence of $\Delta\lambda(\mathbf{m})$ defined in Eq. \ref{eq:dlamdba}. Fig. \ref{fig:smooth_pn}(b) shows the gate voltage dependence of $\Delta\lambda(\mathbf{m})$ and $p(\mathbf{m})$ (defined in Eq. \ref{eq:polarization_TI}) with $\mathbf{m}$ oriented along $\pm\hat{y}$. $\Delta\lambda(\mathbf{m})$ first goes down as we move from PP to PI (I: intrinsic), then goes up a bit and saturates in the NP region. The decrease of
$\Delta\lambda(\mathbf{m})$ in the PP region is due to a mismatch of modes between the gate side and the probe side as the Fermi energy approaches the Dirac point (intrinsic doping) on the gate side. When $V_g=0\;\mathrm{V}$ the Fermi level on the gate side lies exactly on the Dirac point with zero density of states and thus $\Delta\lambda(\mathbf{m})=0$. It is worth mentioning that the `zero' is an idealized simplification. A rigorous calculation involves integration over the bias window which would result in a small but non-zero value. 

When the gate side is switched to the N region, the angular filtering effect shows up and results in a smaller value of $\Delta\lambda(\mathbf{m})$ compared to its symmetric point (with the same $|V_g|$) in the PP region. Since the normal incident mode is not affected by the potential barrier, a small but  near constant $\Delta\lambda(\mathbf{m})$ shows up in the NP region as $V_g$ increases. This asymmetry between PP and NP region and the non-vanishing $\Delta\lambda(\mathbf{m})$ in the NP region separates the TI surface from other 2D systems such as graphene or Rashba systems where there is either $\Delta\lambda(\mathbf{m})=0$ in all regions due to spin degeneracy (graphene) or $\Delta\lambda(\mathbf{m})=0$ in the transmitted N region due to decaying waves in a potential barrier for massive tunneling electrons (Rashba).

We can further demonstrate collimation in TIPNJ by plotting polarization $p(\mathbf{-\hat{y}})$ as a function of the gate voltage, as shown in Fig. \ref{fig:smooth_pn}(b). Electrons moving along the $\mathbf{\hat{x}}$ direction carry $\mathbf{-\hat{y}}$ spin. Right across the NP junction, filtered electrons have a narrower $\mathbf{k}$ distribution compared to the homogeneous PP case, and thus higher (close to $100\%$) spin polarization. In reality, this kind of measurement is limited by the sensitivity of the FM probe, but a clear and significant increase of polarization should be observable as we proceed from homogeneous PP case to NP doping with reasonable $P_\mathrm{FM}$ values. 

\begin{figure}
\includegraphics[width=8cm]{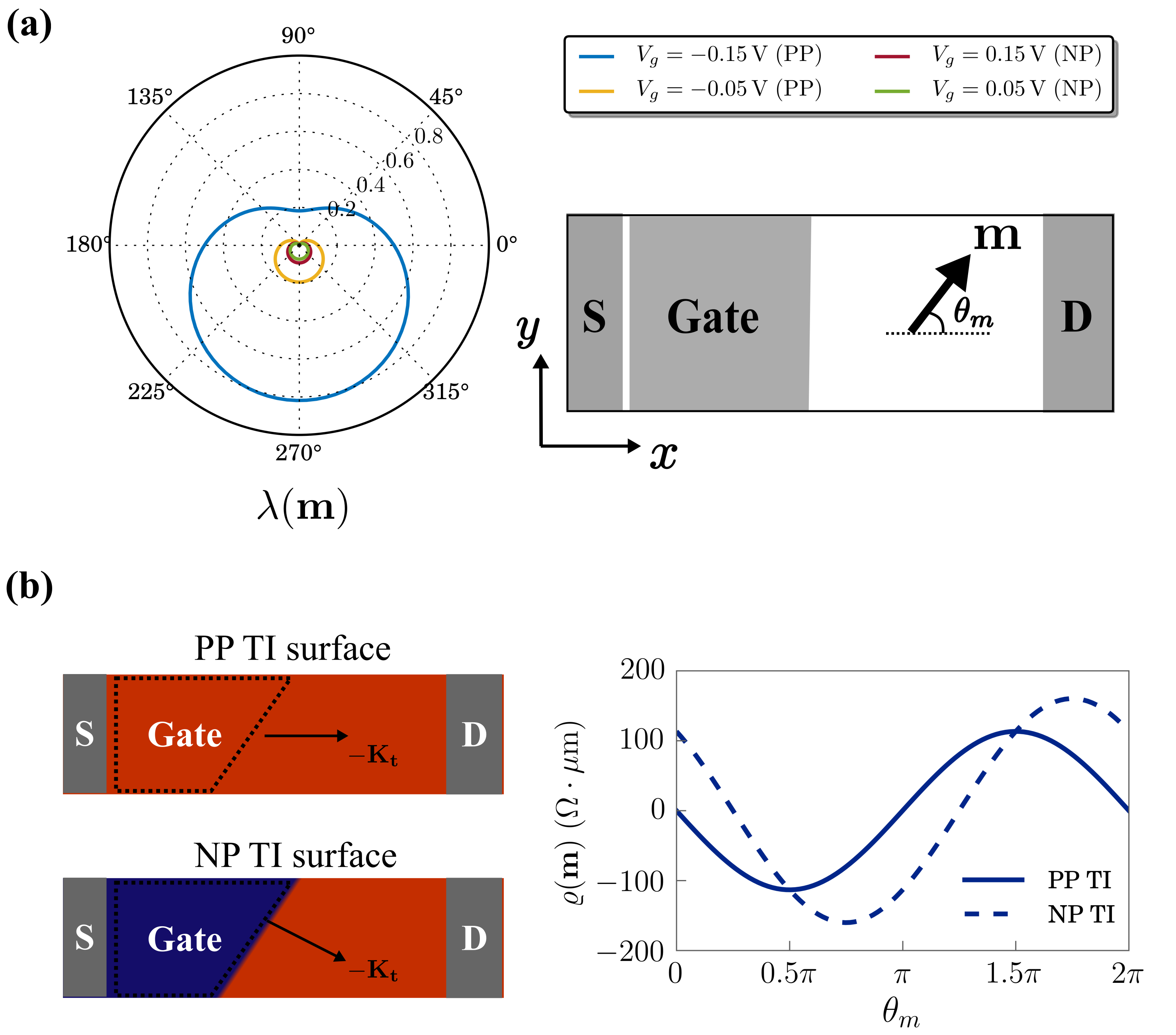}
\caption{(a) Angular dependence of  $\lambda(\mathbf{\hat{m}})$ for different gate voltages. (b) Left. Schematics of a tilted gate on TI surface. (b) Right. Compare the angular function $\mathcal{\varrho}(\mathbf{m})$ in PP and NP case.}
\label{fig:polarpp}
\end{figure}

{\it{Angular dependence of $\lambda(\mathbf{m})$.}}
Our discussion so far focused on measurement along two opposite directions ($\pm\mathbf{\hat{y}}$), assumed to be orthogonal to the electron transport direction. For an arbitrary orientation of the magnetization $\mathbf{m}$, $\lambda(\mathbf{m})$ is a cosine function of the relative angle between the magnetization $\mathbf{m}$ and the spin orientation of the non-equilibrium electrons. Fig. \ref{fig:polarpp}(a) shows the angular dependence of $\lambda(\mathbf{m})$ with different gate voltages. From homogeneous PP to NP junction, apart from the change in the magnitude, $\lambda(\mathbf{m})$ remains the same cosine function. This is because the FM probe cannot isolate individual modes but measures the sum over all transport modes (see \ref{eq:lambda_analy}). In our basic setup, the PN junction filters electrons with large incident angles but the transmitted modes are still symmetrically distributed with respect to $\mathbf{\hat{x}}$. Therefore the average momenta in the PP  and NP junction only differ from each other by their magnitude. To experimentally observe the normal tunneling mode, we can put a tilted gate that is not orthogonal to the transport direction (see Fig. \ref{fig:polarpp}(b)). A tilted gate will not affect the result from the homogeneous case but will collimate the electrons to a different angle for NP, thereby creating a phase shift in the angular dependence of $\lambda(\mathbf{m})$. Since we only care about the phase of $\lambda(\mathbf{m})$, we can define an angular function as:
\begin{eqnarray}
\mathcal{\varrho}(\mathbf{m})&=&\frac{\mu_p(\mathbf{m})-\mu_p(-\mathbf{m})}{qJP_\mathrm{FM}}
\end{eqnarray}
which will scale $\lambda(\mathbf{m})(\mathrm{or}\;\mu_p(\mathbf{m}))$ by the charge current density $J$ and make the PP and NP cases easier to compare, as shown in Fig. \ref{fig:polarpp}(b).

\begin{figure}
\includegraphics[width=8cm]{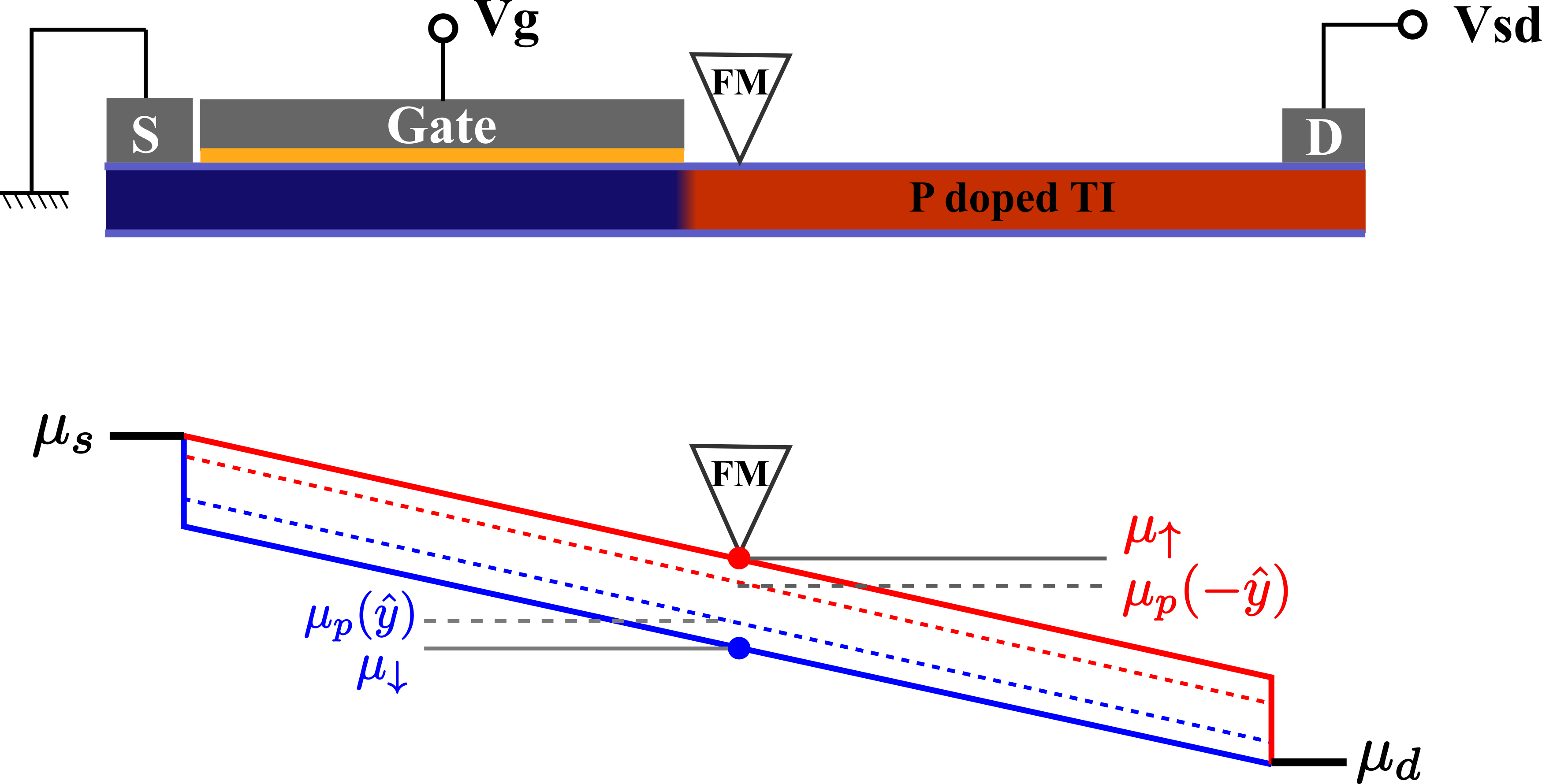}
\caption{Chemical potential profile in a diffusive system.}
\label{fig:diffusive}
\end{figure}

Note that we formulated our equations Eq.\ref{eq:NEGF}-\ref{eq:polarization_TI} assuming a ballistic channel where $\mu_p(\mathbf{m})$ can be directly related to the chemical potentials from the source and drain. However, our analysis in Appendix \ref{sec:analytical} can be easily adopted to a diffusive system with a different interpretation. $\mu_s$ and $\mu_d$ in the previous discussions should be replaced by the local chemical potential $\mu_\uparrow$ and $\mu_\downarrow$ for spin up and spin down channels, as indicated in Fig. \ref{fig:diffusive}. All of our previous discussions are still valid given the following conditions: in a diffusive system, a momentum scattering event can disrupt the collimation effect of the NP junction. To be able to detect the Klein tunneling physics of the junction, the probe needs to be placed very close to the junction, preferably within the mean free path of the TI surface electrons ($\sim 120\,\mathrm{nm}$ estimated in $\mathrm{Bi}_2\mathrm{Te}_3$\cite{xiu2011manipulating}). From the discussion of $p(\mathbf{m})$ earlier, we need information on $\mu_d\,(\mathrm{replace\;by}\;\mu_\downarrow)$ at the junction. One way to do this is to use a normal voltage probe to map out the resistance from junction to the drain to extract the slope shown in Fig. \ref{fig:diffusive}, and then estimate the local electrochemical potential from the applied drain bias. 

\begin{figure}
\includegraphics[width=8cm]{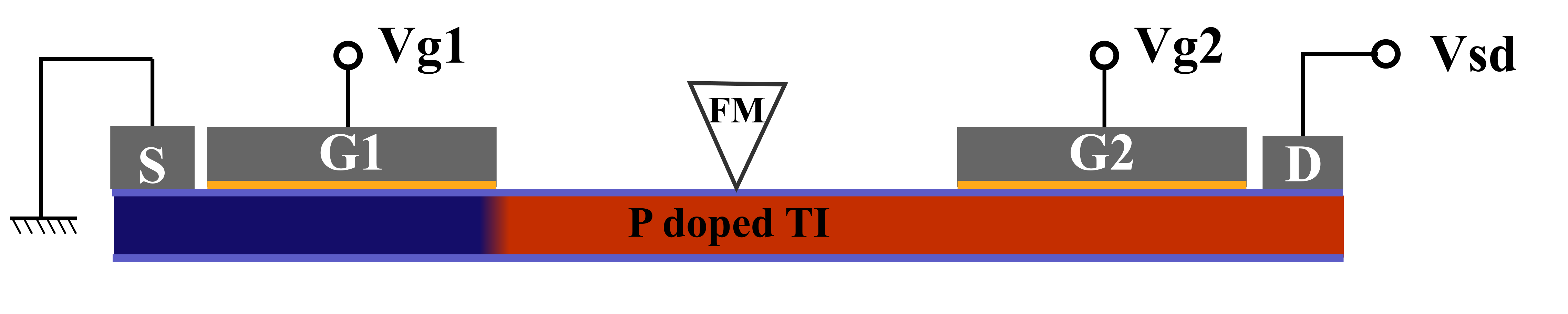}
\caption{One possible experimental measurement set-up. Intrinsically P-doped topological insulator under a N type gate near the source. The FM probe is placed on the exposed P side.}
\label{fig:exp_set_up}
\end{figure}

{\it{Possible experimental set-up.}}
Ideally we would like to rotate the magnetization of the ferromagnetic probe to map out the angle-dependent voltage signals. To our knowledge such a reorientation of an FM probe is challenging. Even fixing the magnetization of the FM probe orthogonal to the transport direction is not straightforward. Instead, we propose placing two separate gates near the source and drain (Fig. \ref{fig:exp_set_up}), creating a symmetric system. Only one of the gates is used at a time to create an N region on one side. When the current direction is switched, we flip the gate polarities on both sides and the entire system is mirrored. Another possibility is to put two probes (one FM, one normal) close to each other and measure the voltage difference between them. It is not difficult to show that $\mu_p(\mathbf{m})-\mu_p(\mathbf{-m})=2(\mu_p(\mathbf{m})-\mu_{nm})$ where $\mu_{nm}$ is the voltage measured at the non-magnetic probe.

%\section{Summary}
To summarize, we propose a straightforward potentiometric measurement on a TIPNJ with a FM probe. Quasi-analytical results are worked out and benchmarked with numerical simulations. Predictable voltage asymmetries features and angular dependences directly bear out signatures of Klein tunneling in the TI, including non-idealities (probe polarization, momentum scattering) that may influence quantitative details seen in the experiment.

%\begin{acknowledgments}
We wish to acknowledge the generous support from NSF Grant No. CCF1514219 and NRI. We are also thankful for the discussions with Prof. Supriyo Datta and his student Shehrin Sayed from Purdue University, Dr. An Ping Lee from Oak Ridge National Laboratory (ORNL), and Prof. Nitin Samarth from Penn State University. This work used Rivanna high performance computing system at the University of Virginia.
%\end{acknowledgments}

%merlin.mbs aipnum4-1.bst 2010-07-25 4.21a (PWD, AO, DPC) hacked
%Control: key (0)
%Control: author (8) initials jnrlst
%Control: editor formatted (1) identically to author
%Control: production of article title (0) allowed
%Control: page (1) range
%Control: year (1) truncated
%Control: production of eprint (0) enabled
%merlin.mbs aipnum4-1.bst 2010-07-25 4.21a (PWD, AO, DPC) hacked
%Control: key (0)
%Control: author (8) initials jnrlst
%Control: editor formatted (1) identically to author
%Control: production of article title (0) allowed
%Control: page (1) range
%Control: year (1) truncated
%Control: production of eprint (0) enabled
%
\clearpage

\appendix

\section{Quasi-analytical derivations for the abrupt PN junction}
\begin{figure}[h]
\includegraphics[width=8cm]{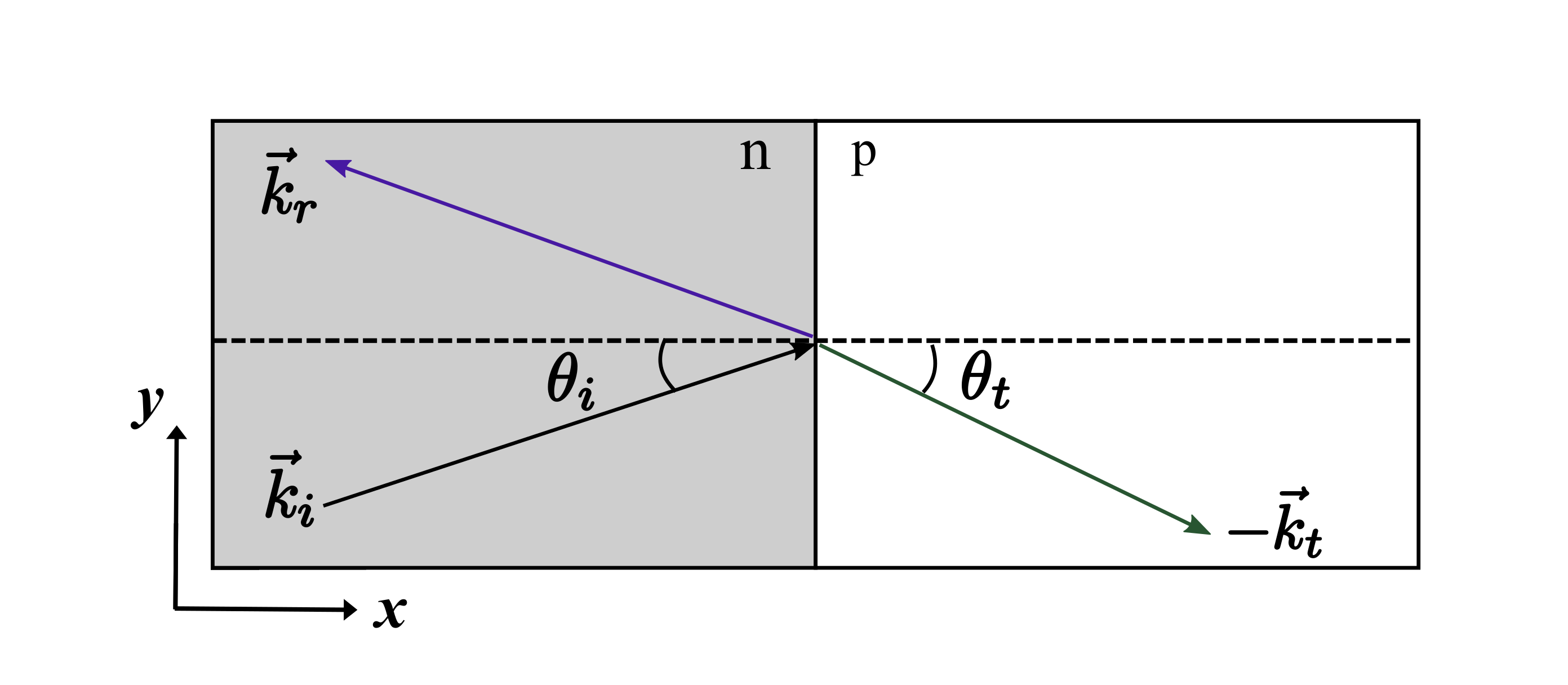}
\caption{Incident, reflected and transmitted electrons waves in a TI pn junction.}
\label{fig:angles}
\end{figure}

From the $\mathbf{k}\cdot\mathbf{p}$ model of TI Hamiltonian, 
\label{sec:analytical}
\begin{equation}
H=v_F\hat{z}\cdot\left(\boldsymbol{\sigma}\times \mathbf{p}\right)=\hbar v_F\left(\sigma^xk_y-\sigma^yk_x\right)
\label{eq:hamil}
\end{equation}
with eigen-wave functions given by:

\begin{eqnarray}
|\psi\rangle_\sigma=\frac{1}{\sqrt{2S}}&\begin{pmatrix} 1 \\ -sie^{i\theta} \end{pmatrix}e^{i\mathbf{k}\cdot\mathbf{r}} \nonumber \\
s=&\mathrm{sgn}(E_\mathbf{k})
\end{eqnarray}
where $\mathrm{sgn}(E_k)=1$ is for the N type and $\mathrm{sgn}(E_k)=-1$ for the P type. The electron wave function is scattered at the junction interface which can be described the set of equations:
\begin{equation}
\begin{split}
|\psi\rangle_\sigma&=|\psi_i\rangle_\sigma+r|\psi_r\rangle_\sigma\\
|\psi\rangle_\sigma&=t|\psi_t\rangle_\sigma
\end{split}
\label{eq:scatter}
\end{equation}
where $|\psi_i\rangle_\sigma,|\psi_r\rangle_\sigma,|\psi_t\rangle_\sigma$ are the incoming, reflected and transmitted electron wave functions respectively (see Fig. \ref{fig:angles}) and $r/t$ is the reflection/transmission coefficient. $\sigma=\uparrow,\downarrow$ is the spin index. Substitute the eigen functions into Eq. \ref{eq:scatter} and match the wave functions at $\mathbf{r}=0$ one gets the transmission coefficient $t$ across the junction from the source to the drain:
\begin{eqnarray}
&\text{for NP}\qquad t=\frac{e^{i\theta_i}-e^{-i\theta_t}}{e^{-i\theta_i}-e^{i\theta_t}} \\ 
&\text{for PP}\qquad t=\frac{e^{i\theta_i}-e^{i\theta_t}}{e^{-i\theta_i}-e^{-i\theta_t}}
\label{eq:transmission}
\end{eqnarray}
It is convenient to replace $\theta_t$ with $\theta_t+\pi$ in the NP case so that the expressions for $t$ are the same in both cases. The relation between the incident angle $\theta_i$ and the transmitted angle $\theta_t$ are given by the conservation of the $y$ component of the wave vector given by $(E-qV_n)\sin\theta_i=(E-qV_p)\sin\theta_t$. In the small bias window near $E_f$, the charge current can be expressed as:
\begin{eqnarray}
I&=&\frac{q}{h}T(E_f)(\mu_s-\mu_d)\nonumber \\
T(E_f)&=&\frac{qV_pW}{hv_F}\int_{-\pi/2}^{\pi/2}|t|^2\cos\theta_td\theta_i
\label{eq:charge_current}
\end{eqnarray}

Knowing the transmission coefficient allows us to calculate $\mathrm{Tr}\left[\Gamma_\mathrm{FM}A_s\right]$:
\begin{eqnarray}
&&\mathrm{Tr}\left[\Gamma_\mathrm{FM}A_s\right]\nonumber \\
&=&W\sum_{v_x(\mathbf{k}_t)>0}[1+P_\mathrm{FM}\mathbf{m}\cdot \mathbf{s}(\mathbf{k}_t)t(\mathbf{k}_t)]\delta(E_f-E(\mathbf{k}_t))\nonumber \\
&=&\sum_{v_x(\mathbf{k}_i)>0}[1+P_\mathrm{FM}\mathbf{m}\cdot \mathbf{s}(\mathbf{k}_t)t(\mathbf{k}_t)]\delta(E_f-E(\mathbf{k}_i))\nonumber \\
&=&\frac{S}{4\pi^2\hbar^2v^2_F}\int_{\pi/2}^{-\pi/2}[1+P_\mathrm{FM}\mathbf{m}\cdot \mathbf{s}(\mathbf{k}_t)t(\mathbf{k}_t)]\delta(E_f-E(\mathbf{k}_t))\nonumber
\label{eq:As_analy}
\end{eqnarray}

To solve $\lambda(\mathbf{m})$ from Eq. \ref{eq:fp} analytically, sum over all modes with positive group velocity:
\begin{eqnarray}
&&\lambda(\mathbf{m}) = \frac{ \mathrm{Tr}\left[\Gamma_\mathrm{FM}A_s\right]}{\mathrm{Tr}\left[\Gamma_\mathrm{FM}A\right]}\nonumber \\
&=&\frac{\sum_{v_x(\mathbf{k}_t)>0}[1+P_\mathrm{FM}\mathbf{m}\cdot \mathbf{s}(\mathbf{k}_t)t(\mathbf{k}_t)]\delta(E_f-E(\mathbf{k}_t))}{\sum_{\mathbf{k}_t}[1+P_\mathrm{FM}\mathbf{m}\cdot \mathbf{s}(\mathbf{k}_t)]\delta(E_f-E(\mathbf{k}_t))}\nonumber \\
&=&\frac{\sum_{v_x(\mathbf{k}_t)>0}[1+P_\mathrm{FM}\mathbf{m}\cdot \mathbf{s}(\mathbf{k}_t)t(\mathbf{k}_t)]\delta(E_f-E(\mathbf{k}_t))}{\sum_{\mathbf{k}_t}\delta(E_f-E(\mathbf{k}_t))}\nonumber\\
\label{eq:fp_analy}
\end{eqnarray}
with $t(\mathbf{k}_t)=|t|^2$ given in Eq. \ref{eq:transmission}. The last part of Eq. \ref{eq:fp_analy} holds because each pair of states $\mathbf{k}_t,-\mathbf{k}_t$ cancel each other due to the time reversal symmetry of TI surface Hamiltonian $\mathbf{s}(\mathbf{k}_t)=-\mathbf{s}(-\mathbf{k}_t)$.
Assume the ferro-magnetic voltage probe has an in-plane magnetization $(m_x,m_y)$. Substitute the transmission coefficient into Eq. \ref{eq:fp_analy} and replace $\sum$ with $\frac{S}{4\pi^2}\int d^2k$. For the numerator, it is easier to sum the modes from the incoming side ($\sum_{\mathbf{k}_t}\Rightarrow\sum_{\mathbf{k}_i}$). For the denominator, notice it is just the density of states on the P side. Therefore $\lambda(\mathbf{m})$ can be written as:
\begin{eqnarray}
&&\lambda(\mathbf{m})=
\left|\frac{E_f-qV_g}{E_f-qV_p}\right|\times\nonumber\\
&&\int_{-\pi/2}^{\pi/2}\frac{\cos^2\theta_i\left(1 +P_\mathrm{FM}m_x\sin\theta_t-P_\mathrm{FM}m_y\cos\theta_t\right)}{2\pi\cos^2\left(\frac{\theta_i+\theta_t}{2}\right)}d\theta_i \nonumber\\
\label{eq:lambda_analy}
\end{eqnarray}
where $\theta_t=\sin^{-1}[(E-qV_g)/(E-qV_p)\sin\theta_i^L]$. In the case of homogeneous PP TI surface, $V_g=V_p$ and $\theta_i=\theta_t$. Eq. \ref{eq:lambda_analy} reduces to:
\begin{eqnarray}
\lambda(\mathbf{m})&=&\int_{-\pi/2}^{\pi/2}\frac{\left(1 +P_\mathrm{FM}m_x\sin\theta_i-P_\mathrm{FM}m_y\cos\theta_i\right)}{2\pi}d\theta_i \nonumber \\
&=&\frac{\pi-2P_\mathrm{FM}\mathbf{m}\cdot\mathbf{\hat{y}}}{2\pi}
\label{eq:lambda_PP}
\end{eqnarray}
We leave the discussion of symmetric NP junction case in the next section with smooth transition between N region and P region. 

\section{Extension to the smooth PN junction}
Compare to the abrupt junction, a smooth NP junction with linear transition depicted in Fig. \ref{fig:TI_struc}(b) adds an exponential factor to the transmission coefficient calculated from the abrupt junction $|t|^2$ (Eq.\ref{eq:transmission}). Here we can directly borrow the result for the transmission coefficient from graphene PN junction\cite{sajjad2013manipulating}:

\begin{equation}
\begin{split}
|t|^2_{\mathrm{smooth}}&=\frac{\cos^2\theta_i}{\cos\left(\frac{\theta_i+\theta_t}{2}\right)}e^{-\pi\frac{k_ik_t}{k_i+k_t}\sin\theta_i\sin\theta_td}\\
&=\frac{\cos^2\theta_i}{\cos\left(\frac{\theta_i+\theta_t}{2}\right)}e^{-\pi\frac{\hbar v_F}{qV_0} k_t^2\sin^2\theta_td}
\end{split}
\label{eq:smooth_pn}
\end{equation}
where $d$ is the transition length between N region and P region. $V_0=|V_p-V_n|$ is the potential difference from N region to P region. By replacing $|t|^2$ with $|t|^2_\mathrm{smooth}$ one can extend the result of $\lambda(\mathbf{m})$ from the abrupt junction. For symmetric NP junction with linear transition $V_g=-V_p$ and $\theta_t=-\theta_i$. We can evaluate $\lambda(\mathbf{m})$ at $E_F=0$:
\begin{eqnarray}
\lambda(\mathbf{m})&=&\frac{1}{2\pi}\int_{-\pi/2}^{\pi/2}\cos^2\theta_ie^{-\frac{\pi qV_nd}{2\hbar v_F}\sin^2\theta_i}d\theta_i \nonumber \\
-&&\frac{P_\mathrm{FM}\mathbf{m}\cdot\mathbf{\hat{y}}}{2\pi}\int_{-\pi/2}^{\pi/2}\cos^3\theta_ie^{-\frac{\pi qV_nd}{2\hbar v_F}\sin^2\theta_i}d\theta_i
\label{eq:lambda_NP}
\end{eqnarray}

\section{Numerical methods}
\label{sec:numerical}
The numerical results are generated from the Non-Equilibrium Green's Function (NEGF) method. An artificial term $\sigma^z=\gamma\hbar v_F\sigma^z(k_x^2+k_y^2)$ is added the avoid the fermion doubling problem as previous studies\cite{hong2012modeling,habib2015chiral}. The TI surface Hamiltonian Eq.~\ref{eq:TI_H} is discretized on a square lattice by the finite difference method\cite{habib2015chiral}:
\begin{eqnarray}
    H&=&\sum_i\epsilon c_i^\dagger c_i + \sum_i\left(t_xc^\dagger_{i,i}c_{i,i+1}+\mathrm{H.C.}\right)\nonumber\\
    &&\qquad\qquad+\sum_j(t_yc^\dagger_{j,j}c_{j,j+1}+\mathrm{H.C.}) \\
    \epsilon&=&-4\hbar v_F\frac{\alpha}{a}\sigma^z\\
    t_x&=&\hbar v_F\left[\frac{i}{2a}\sigma^y+\frac{\alpha}{a}\sigma^z\right]\\
    t_y&=&\hbar v_F\left[-\frac{i}{2a}\sigma^x+\frac{\alpha}{a}\sigma^z\right]
\end{eqnarray}
where $a$ is the square mesh size ($a=5\,\mathrm{nm}$ is chose for the simulations). $\alpha=\gamma/a$ is a fitting parameter and $\alpha=1$ describes the correct bandstructure near the Dirac cone\cite{habib2015chiral}. Periodic boundary condition is assumed in the transverse direction to simulate infinitely wide TI surface. The retarded green's function is given by:
\begin{equation}
G^R(E,\mathbf{k})=(E+\delta - H(\mathbf{k})-\Sigma_s(E,\mathbf{k})-\Sigma_d(E,\mathbf{k}))^{-1}
\end{equation}
where $E$ is the energy and $\mathbf{k}$ is the transverse wavevector. The FM probe is assumed to be weakly coupled to the TI surface so the effect of $\Sigma_p$ on electron transport is neglected when calculating $G^R(E,\mathbf{k})$. Following Eq.~\ref{eq:NEGF}-\ref{eq:couple} and sum over $\mathbf{k}$ for $A_s(\mathbf{k}),A(\mathbf{k})$ in Eq.~\ref{eq:fp} would give $f_p(m)$. 

\section{Angular dependence for tilted junction}
Here we modify Eq.\ref{eq:lambda_analy} for tilted junction shown in Fig.\ref{fig:polarpp} and relate it to the experimental measurable quantities $\mu_p$ and $I$. Denote $\mathbf{G}$ as the normal vector to the junction interface. From Eq.\ref{eq:mup} we have:
\begin{eqnarray}
\mu_p(\mathbf{m})-\mu_p(-\mathbf{m})=(\lambda(\mathbf{m})-\lambda(-\mathbf{m}))(\mu_s-\mu_d)
\end{eqnarray}
where $\Delta\lambda(\mathbf{m})=\lambda(\mathbf{m})-\lambda(-\mathbf{m})$ is given by:
\begin{eqnarray}
&&\Delta\lambda(\mathbf{m})\nonumber \\
&=&\frac{2P_\mathrm{FM}\sum_{v_x(\mathbf{k}_t)>0}\mathbf{m}\cdot \mathbf{s}(\mathbf{k}_t)t(\mathbf{k}_t)\delta(E_f-E(\mathbf{k}_t))}{\sum_{\mathbf{k}_t}\delta(E_f-E(\mathbf{k}_t))} \nonumber \\
&=&\frac{P_\mathrm{FM}\mathbf{m}\cdot \mathbf{S}}{\pi}\\
&&\mathbf{S}=\sum_{v_x(\mathbf{k}_t)>0}\mathbf{s}(\mathbf{k}_t)t(\mathbf{k}_t)
\label{eq:delta_lambda}
\end{eqnarray}
We can rewrite the transmission of the junction from \ref{eq:charge_current} as:
\begin{eqnarray}
T(E_f)&=&\frac{qV_pW}{h v_F}\sum_{v_x(\mathbf{k}_t)>0}\mathbf{\hat{x}}\cdot\mathbf{\hat{k}}_tt(\mathbf{k}_t)\nonumber\\
&=&\frac{qV_pW}{hv_F}\mathbf{\hat{x}}\cdot\mathbf{K}\\
\mathbf{K}&=&\sum_{v_x(\mathbf{k}_t)>0}\mathbf{\hat{k}}_tt(\mathbf{k}_t)\nonumber\\
J&=&\frac{q}{Wh}T(E_f)(\mu_s-\mu_d)\nonumber
\label{eq:current}
\end{eqnarray}
where $\mathbf{\hat{k}_t}$ is the unit vector along momentum $\mathbf{k}_t$. It is easy to see $\mathbf{S}=\mathbf{\hat{z}}\times\mathbf{K}$ due to the spin momentum locking. Therefore the quantity we define can be expressed as:
\begin{eqnarray}
\mathcal{\varrho}(\mathbf{m})&=&\frac{\mu_p(\mathbf{m})-\mu_p(-\mathbf{m})}{qJP_\mathrm{FM}}\nonumber \\
&=&\frac{h^2v_F}{\pi q^3V_p}\frac{\mathbf{m}\cdot \mathbf{S}}{\mathbf{\hat{x}}\cdot\mathbf{K}}\nonumber \\
&=&\frac{h^2v_F}{\pi q^3V_p}\frac{(\mathbf{m}\times\mathbf{\hat{z}})\cdot\mathbf{K}}{\mathbf{\hat{x}}\cdot\mathbf{K}}
\end{eqnarray}
For a homogeneous PP junction, $\mathbf{K}\propto \mathbf{-\hat{x}}$ and $\mathbf{S}\propto \mathbf{-\hat{y}}$. For NP case, only normal mode can pass through the junction, which means $\mathbf{K}\propto \mathbf{-G}$. The change of direction for $\mathbf{K}$ would cause a phase difference for the angular dependence of $\mathcal{\varrho}(\mathbf{m})$.

\end{document}